\documentclass[
nopreprint															
]{jasatex}
\usepackage{amsfonts}
\usepackage{amssymb}
\usepackage{amsmath}
\usepackage{color}
\usepackage{graphicx}
\usepackage{bm}
\usepackage{fancybox}
\usepackage{subfigure}

\newlength{\figtotalheight}
\newcommand{\beq}{\begin{equation}}     \newcommand{\eeq}{\end{equation}}  
\newcommand{\ber}{\begin{eqnarray}}     \newcommand{\eer}{\end{eqnarray}}
\newcommand{\bse}{\begin{subequations}} \newcommand{\ese}{\end{subequations}}

\def\bd#1{\mbox{\boldmath$\displaystyle\mathbf{#1}$} }
\def\dd{\operatorname{d}} 
\def\tr{\operatorname{tr}} 
\def\div{\operatorname{div}} 

\newcommand{\pdd}[2]{\frac{\partial #1}{\partial #2}}

\begin{document}

\title{Special transformations for pentamode acoustic cloaking\footnote{Special issue of JASA: Acoustic Metamaterials}}

\author{Nachiket H. Gokhale}
\email{gokhale@wai.com}
\affiliation{Weidlinger Associates Inc, 375 Hudson Street, New York, NY 10014}
\author{Andrew N. Norris}
\affiliation{Mechanical and Aerospace Engineering, Rutgers University, Piscataway, NJ 08854}
\author{Jeffrey L. Cipolla}
\affiliation{Weidlinger Associates Inc, 375 Hudson Street, New York, NY 10014}

\date{\today}
\pacs{43.20Mv, 43.40.Sk, 43.30.Wi, 43.20.El} 
\keywords{acoustic cloaking, metamaterial, pentamode, transformation acoustics}

\begin{abstract}
The acoustic cloaking theory of Norris\cite{Norris08b} permits considerable freedom in choosing the transformation $f$ from physical to virtual space. The standard process for defining cloak materials is to first define $f$ and then evaluate whether the materials are practically realizable. In this paper, this process is inverted by defining desirable material properties and then deriving the appropriate transformations which guarantee the cloaking effect. Transformations are derived which result in acoustic cloaks with special properties such as  1) constant density 2) constant radial stiffness 3) constant tangential stiffness 4) power-law density 5) power-law radial stiffness 6) power-law tangential stiffness. 7) minimal elastic anisotropy.  
\end{abstract}

\maketitle

\section{Introduction}\label{sec1}

\textit{Acoustic cloaking} refers to making an object invisible to sound waves. This is achieved by enclosing the object of interest with an \textit{acoustic cloak} which guides waves around the object. The cloak leaves the wave-field outside the cloak indistinguishable from the wave-field without the object present. The phenomenon of cloaking is not restricted to acoustics but can occur for different types of waves such as electromagnetic waves \cite{Pendry08a}, elastic waves \cite{Norris11a}, and in a more exotic example, quantum mechanical systems \cite{zhang:matterwaves}.  We restrict our attention here to acoustic cloaking, and specifically pentamode acoustic cloaking for which the density is isotropic. The reader is referred to the review articles by Bryan and Leise\cite{bryanleise:siaminvis} and Greenleaf \textit{et al.}\cite{GreenleafSiamReview} for a comprehensive review of different types of cloaking, its historical development and relation to previous work in inverse problems \cite{Greenleaf03b}.  A review dedicated to acoustic cloaking and transformation acoustics is provided by Chen and Chan \cite{Chen10a}. 

Initial work in acoustic cloaking \cite{Cummer07,Chen07,Cummer08} was based on  transformation optics as developed  by Pendry \textit{et al.} \cite{Pendry06}. Cummer \textit{et al.} \cite{Cummer07} mapped the $\text{2D}$  acoustic equations in a fluid to the single polarization Maxwell's equations, while Chen \textit{et al.}\cite{Chen07} mapped the $\text{3D}$ acoustic equation to the direct current conductivity equation in $\text{3D}$. Cummer \textit{et al.}\cite{Cummer08} derived a formulation for $\text{3D}$ acoustic cloaking starting from scattering theory. These formulations achieved acoustic cloaking using anisotropic density and isotropic stiffness. Norris \cite{Norris08b} provided a formulation of acoustic cloaking which using both anisotropic inertia and stiffness, and as a special case, derived a formulation using isotropic density and anisotropic stiffness. 

\begin{figure}
\centering
\includegraphics[totalheight=6cm]{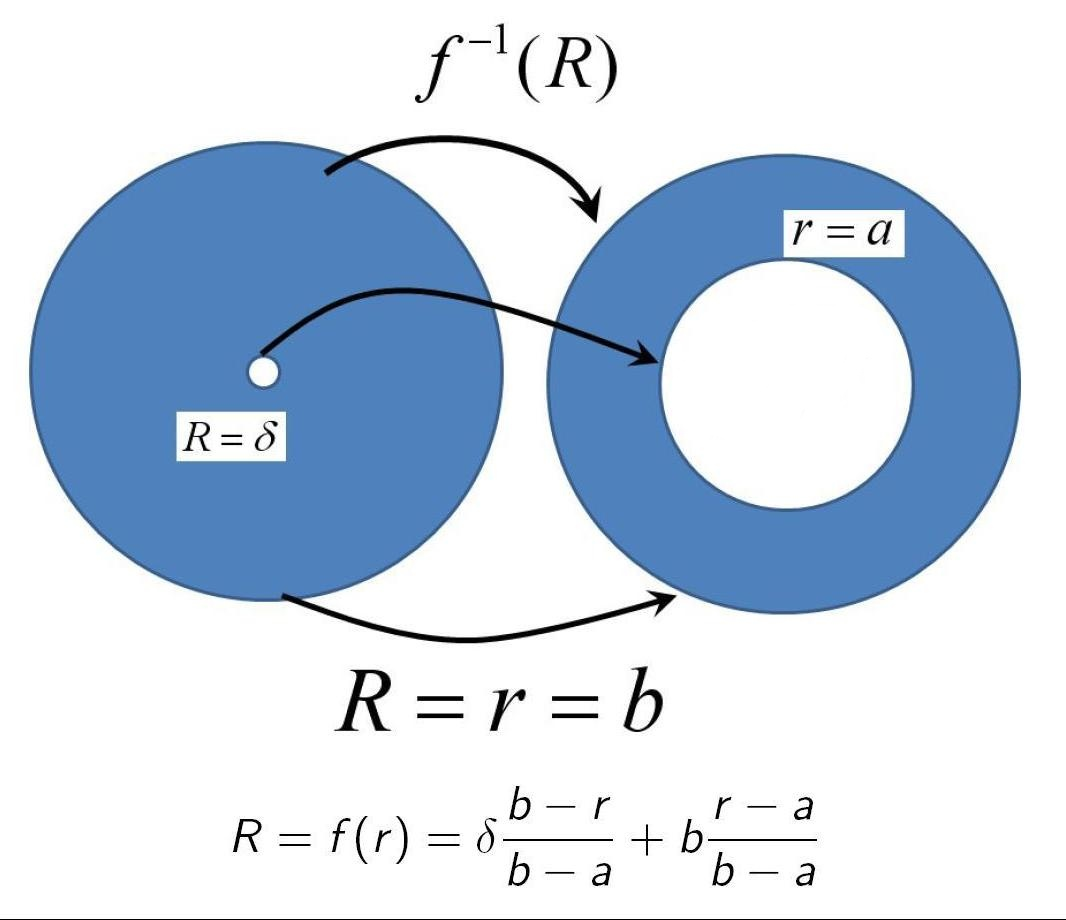}
\caption{\label{norris:transformation} The transformation $f$ from physical space (right) to virtual space (left). The particular mapping shown is the linear mapping of eq.\ (\ref{eqn:threetransformations}a).}
\end{figure}

The acoustic cloaking theory of Norris\cite{Norris08b} involves mapping the physical space to the virtual space using the transformation $f$ as illustrated in Figure \ref{norris:transformation}. The material properties of the cloak can be obtained by choosing the transformation $f$ and using eqs.\ (\ref{eqn:pentak}) to compute the material properties. In practice, however, the properties obtained may not be useful because they are unattainable. In this paper, we derive special forms of $f$ which may result in physically realizable cloaking \textit{metamaterials}, which are composite materials whose macroscopic acoustic properties are controlled by engineering their microstructure. The design and fabrication of such acoustic metamaterials is possible because of recent advances in material science and engineering.  Cloaking metamaterials may have spatially varying anisotropic density and stiffness.  We  restrict our attention here to spatially varying material properties with isotropic density and anisotropic stiffness because Norris\cite{Norris08b} showed that anisotropic density implies that the acoustic cloak has infinite mass. He presented an alternative acoustic cloaking formulation involving \textit{pentamode materials} which have isotropic density and a special type of anisotropic stiffness. Since cloaking is achieved with anisotropic stiffness as opposed to density, it is expected to have frequency independent behavior in theory. 
In practice, however, the behavior is expected to be only extremely wideband or weakly frequency dependent because of frequency limitations arising from:
\begin{itemize} 
\item{ the size of the virtual cloak radius $\delta$ as compared to the wave-length of the incident acoustic wave $\lambda$. Cloaking is ineffective for incident acoustic waves whose wavelength is of the same order as the virtual cloak radius $\delta$.} 
\item{ the length scale of periodic structures present in the composite material used to fabricate the pentamode material.}
\item{ an intrinsic frequency dependence in the properties of the composite material. }
\end{itemize}

The paper is organized as follows. Section \ref{sect:pentamodetheory} provides a short review of the pentamode acoustic cloaking theory\cite{Norris08b}. Next, in \S\ref{sect:constantspatial}, we derive transformations $f$ which yield specialized spatial distributions of material properties, namely,  1) constant density 2) constant radial stiffness 3) constant tangential stiffness  and explain the wave-propagation with ray-tracing.   Such distributions may be simple to manufacture and may also help in evaluating the feasibility of manufacturing material properties on Ashby charts, as in\cite{smithurzhumov:attainable-njp}. We note that in related work, Cummer\cite{cummer:nonsingular-constants} has derived transformations for electromagnetic cloaking which yield constant magnetic permittivity $\epsilon_\phi$. In \S\ref{sect:powerlaw}  we derive transformations which yield 1) power-law density 2) power-law radial stiffness 3) power-law tangential stiffness. In \S\ref{sect:optimalaniso}, we derive a distribution of elastic properties that minimizes the elastic anisotropy. 

\section{Review of acoustic cloaking using pentamode materials}
\label{sect:pentamodetheory} 

Acoustic cloaking relies on a transformation from an undeformed or original domain $\Omega$ to a current (deformed) domain $\omega$ which is given by the point-wise deformation 
${\bd X}\in \Omega$ $\rightarrow$   ${\bd x} \in \omega$.   
Using notation from the theory of finite elasticity, the deformation gradient is defined 
${\bd F} = \nabla_X {\bd x}$, 
or in component form $F_{iI}  = \partial x_i /\partial X_I$.    
The Jacobian of the deformation is $J = \det {\pmb F}$,  or in terms of volume elements in the two 
configurations, $J = \dd v/\dd V$. The polar decomposition is ${\pmb F} = {\pmb V} {\pmb R} $, where ${\pmb R}$
 is proper  orthogonal ($ {\pmb R}{\pmb R}^t = {\pmb R}^t {\pmb R} = {\pmb I}$, $\det {\pmb R}= 1$) and the left 
stretch tensor ${\pmb V}$ is the  positive definite solution of 
${\pmb V}^2 = {\pmb B}$ where ${\pmb B}$ 
is the left Cauchy-Green or Finger  tensor 
${\pmb B}  = {\pmb F}{\pmb F}^t$. 

For a given transformation the cloaking material is not unique\cite{Norris08b}.  For instance, the inertial cloak \cite[eq.\ (2.8)]{Norris08b} is defined by the density tensor ${\pmb \rho} =\rho_0 J{\bd B}^{-1}$ and  bulk modulus $K = K_0J$.  At the other  end of the spectrum of possible materials is the pentamode cloak with isotropic density, which can be chosen if the deformation satisfies the property\cite[Lemma 4.3]{Norris08b} that there is  a function $h({\bd x})$ for which   $\div h {\bd V} =0$.  This is the case for radially symmetric deformations in $\text{2D}$ and $\text{3D}$, the cylinder and sphere, respectively. The  pentamode material is then\cite[eq.\ (4.8)]{Norris08b} 
$ \rho =\rho_0 J^{-1}$,  $K = K_0J$, ${\bd S}=J^{-1} {\bd V}$, where the fourth order elasticity  tensor is 
${\bd C}= K {\bd S}\otimes {\bd S}$. 

Radially symmetric deformations in $\text{2D}$ and $\text{3D}$ are defined by $R^{-1}{\bd X}= r^{-1}{\bd x}$ where $R=|{\bd X}|$, $r=|{\bd x}|$. If we let $R= f(r)$, then the inverse mapping is defined as ${\bd X} = f(r) r^{-1}{\bd x}$. 
In this case we can identify a ``radial bulk modulus" and an orthogonal bulk modulus
\beq\label{1=1}
K_r \equiv  C_{rrrr}=K_0J^{-1} B_{rr}, 
\quad
K_\perp \equiv K_0J^{-1} B_{\theta\theta}, 
\eeq 
where $\theta$ denotes a direction orthogonal to the radial direction. 

The requirements on the transformation generator $f(r)$ admit an infinity of functions. In this paper, we will take advantage of this fact to design transformations which result in desirable material properties. One familiar transformation, used frequently in acoustic cloaking work, is shown in Figure \ref{norris:transformation}. Some examples from the infinite family of permissible transformations are:
\beq
\label{eqn:threetransformations}
R=f(r) = \begin{cases}
\big( \frac{b-\delta}{b-a}\big)r - b\big( \frac{a-\delta}{b-a}\big), & (a)\\
b\, \big(\frac{r}{b}\big)^{\frac{\ln (b/\delta)}{\ln (b/a)}} , &(b)\\
 \bigg[\bigg( \frac{b^d-\delta^d}{b^d-a^d}\bigg)r^d - \bigg( \frac{a^d-\delta^d}{b^d-a^d}\bigg)b^d\bigg]^\frac{1}{d}. 
&(c)
\end{cases}
\eeq
where $b$, $a$ and $\delta$ are the outer radius of the cloak, the inner radius of the cloak, and the virtual cloak radius, as shown in Figure \ref{norris:transformation}. The linear transformation (\ref{eqn:threetransformations}a) is known as the KSVW 
mapping for the paper \cite{Kohn08}   in which it was first used extensively in this form. Figure \ref{fig:ksvw} shows rays passing through a KSVW cloak. The power law mapping (\ref{eqn:threetransformations}b) yields  constant $\rho_r$ for the inertial cloak  and  constant $K_r$ for the pentamode cloak in $\text{2D}$. As we will see below, the transformation (\ref{eqn:threetransformations}c) yields constant $K$ for the inertial cloak  or constant $\rho$ for the pentamode cloak. 


\section{Transformations yielding constant material property distributions}
\label{sect:constantspatial} 

In this section we determine the transformations $f$ that yield constant spatial distributions of material properties. We consider a constant distribution  of $\rho$ in \S\ref{sect:consrho}, constant $K_r$ in \S\ref{sect:conskr}, and constant $K_\perp$ in \S\ref{sect:conskt}, respectively. Both $\text{2D}$ and $\text{3D}$ cases are considered.  The conditions for feasibility are summarized in Table \ref{tab:constantprops}. 

\protect{
\begin{center}
\begin{table}
\begin{tabular}{|c|c|c|c|}
\hline
Constant & d=2       & d=3  & Conditions other than $\rho_0, K_0 > 0$\\
\hline
$\rho$   & \checkmark    & \checkmark & $\rho_c > \rho_0 $ \\
\hline
$K_r$    & \checkmark   & \checkmark &  Conditional in 3D, $K_r < \frac{K_0\delta}{a}$, $K_r > 0$\\
\hline
$K_\perp$ & \checkmark  & \checkmark & Conditional in 3D, $K_0 < K_\perp$\\
\hline
\end{tabular}
\caption{\label{tab:constantprops} Feasibility of constant material properties in $\text{2D}$ $(d=2)$ and $\text{3D}$ $(d=3)$.}
\end{table}
\end{center}
}
The density $\rho$ (isotropic), the radial stiffness $K_r$, and the tangential stiffness $K_\perp$ for a pentamode cloak surrounded by an anisotropic fluid with density $\rho_0$ and bulk modulus $K_0$  satisfy the following relations:
\beq\label{eqn:pentak}
\begin{aligned}
 &K_r(r) = K_{0}\frac{1}{f'} \Big( \frac{f}{r}\Big)^{d-1}, 
\ \ K_\perp(r) = K_{0}f' \Big( \frac{f}{r}\Big)^{d-3}, 
\\
& \qquad \qquad 
\rho(r) = {\rho_0}f'\Big( \frac{f}{r}\Big)^{d-1},  
\end{aligned}
\eeq
Our procedure to determine $f$ consists of treating  eqs.\ \eqref{eqn:pentak}  as differential equations for $f$ with the material properties ($\rho,K_r,K_\perp$) known. Having determined $f$ we will prove that it satisfies the necessary conditions $f>0 , f'> 0$ for $r  \in [a,b]$, the existence of $b$ such that $f(b)=b$, and the existence of $\delta$ such that $f(a)=\delta << a$. 

We remark that eqs.\ \eqref{eqn:pentak} are consistent with the connection between the three parameters which is independent of the transformation:
\beq\label{3=3}
K_0^{-d}K_rK^{d-1}_\perp = (\rho/\rho_0)^{d-2}.
\eeq
Equations \eqref{eqn:pentak}  also imply 
that at the edge of the cloak, the cloak is \textit{impedance matched} in the radial direction but not in the tangential direction:
\beq
\begin{aligned}
Z_r (r) &\equiv \sqrt{K_r\rho} 
\ \ \Rightarrow \ \ Z_r(b) = \sqrt{K_0\rho_0} \equiv Z_0\, ,
\\
Z_\perp (r) &\equiv \sqrt{K_\perp\rho} 
\ \ \Rightarrow \ \ Z_\perp (b) = Z_0{f'}(b).
\end{aligned}
\eeq 
In contrast, at the edge of the cloak, the wave speeds are matched in the tangential direction, but not in the radial direction:
\beq
\begin{aligned}
c_r (r) &\equiv   \sqrt{{K_r}/{\rho}} 
\ \ \Rightarrow \ \ c_r (b) = {c_0}/{f'}(b) ,
\\ 
c_\perp (r)&\equiv  \sqrt{{K_\perp}/{\rho}} 
\ \ \Rightarrow \ \ c_\perp(b) = c_0 .
\end{aligned}
\eeq

\subsection{\label{sect:consrho}Transformations yielding constant cloak density $\rho$}
We assume that we are given the cloak geometry $b>a>0$, and using equation (\ref{eqn:pentak}) and $\rho(r)=\rho_c$, a constant, we determine $f$ and prove that $f>0,f'>0$ for  $r\in\,(a,b)$. 
Solving (\ref{eqn:pentak}) for $f$ yields 
\beq
\label{eqn:fdefrhocons}
f = \big[ b^d  +\frac{\rho_c}{\rho_0} \big(r^d -b^d \big)
\big]^{1/d},
\eeq
and differentiating implies  
$f' 
> 0 $ for $ \rho_c > 0$, showing that $f$ is monotonically increasing. Since $f(a) = \delta > 0 $, $f>0 \in (a,b)$. Enforcing $\delta=f(a) < a $ yields $\rho_c > \rho_o$. This result makes physical sense because the deformation $f^{-1}$ compresses the volume of fluid into a smaller volume. $K_r, K_\perp$ are determined by using $f$ in equation (\ref{eqn:pentak}) and are given by:
\beq
\label{eqn:kperpdefrhocons}
K_r    = K_{0}\frac{\rho_{0}}{\rho_{c}}\bigg(\frac{f}{r}\bigg)^{2(d-1)}, \ \ K_\perp = K_0 \frac{\rho_c}{\rho_0}\bigg(\frac{r}{f}\bigg)^2.
\eeq
Similarly $\delta$ and its sensitivity can be determined by 
\beq
\label{eqn:deltadefrhocons}
\delta = \big[ b^d  +\frac{\rho_c}{\rho_0} \big(a^d -b^d \big)
\big]^{1/d},
\ \ \pdd{\delta}{\rho_c}  = \frac{a^d - b^d}{\rho_0{d}\delta^{d-1}} {< 0}.
\eeq
Finally, note that, $K_r \propto \delta^{2(d-1)} \text{ and } K_\perp \propto { \delta^{-2}}$ In $\text{3D}$ this implies a very strong decrease in $K_r$ with $\delta$. 

\begin{figure}
\centering
\subfigure[\label{fig:ksvw} KSVW cloak]{
  \includegraphics[totalheight=\figtotalheight]{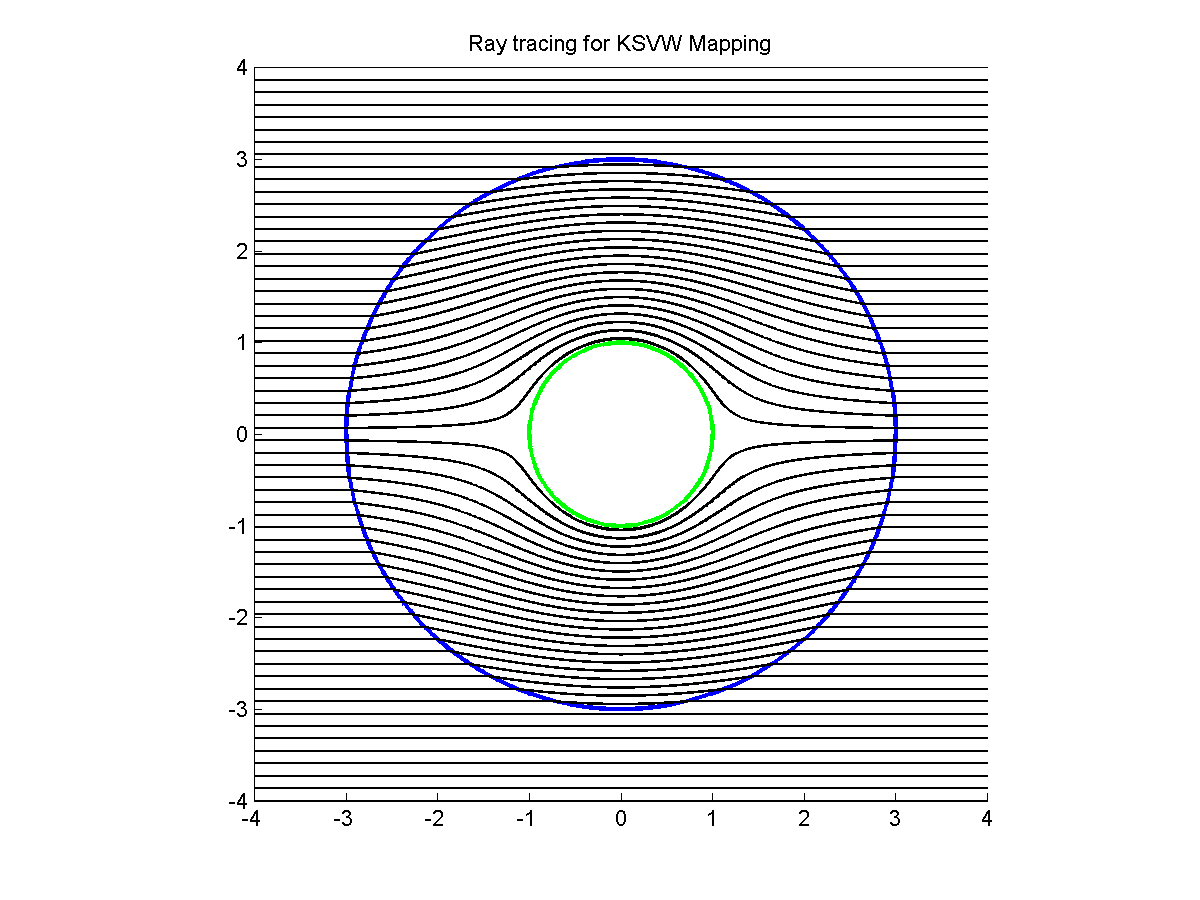}
}
\subfigure[\label{fig:rayrhocons} Constant density cloak]{
  \includegraphics[totalheight=\figtotalheight]{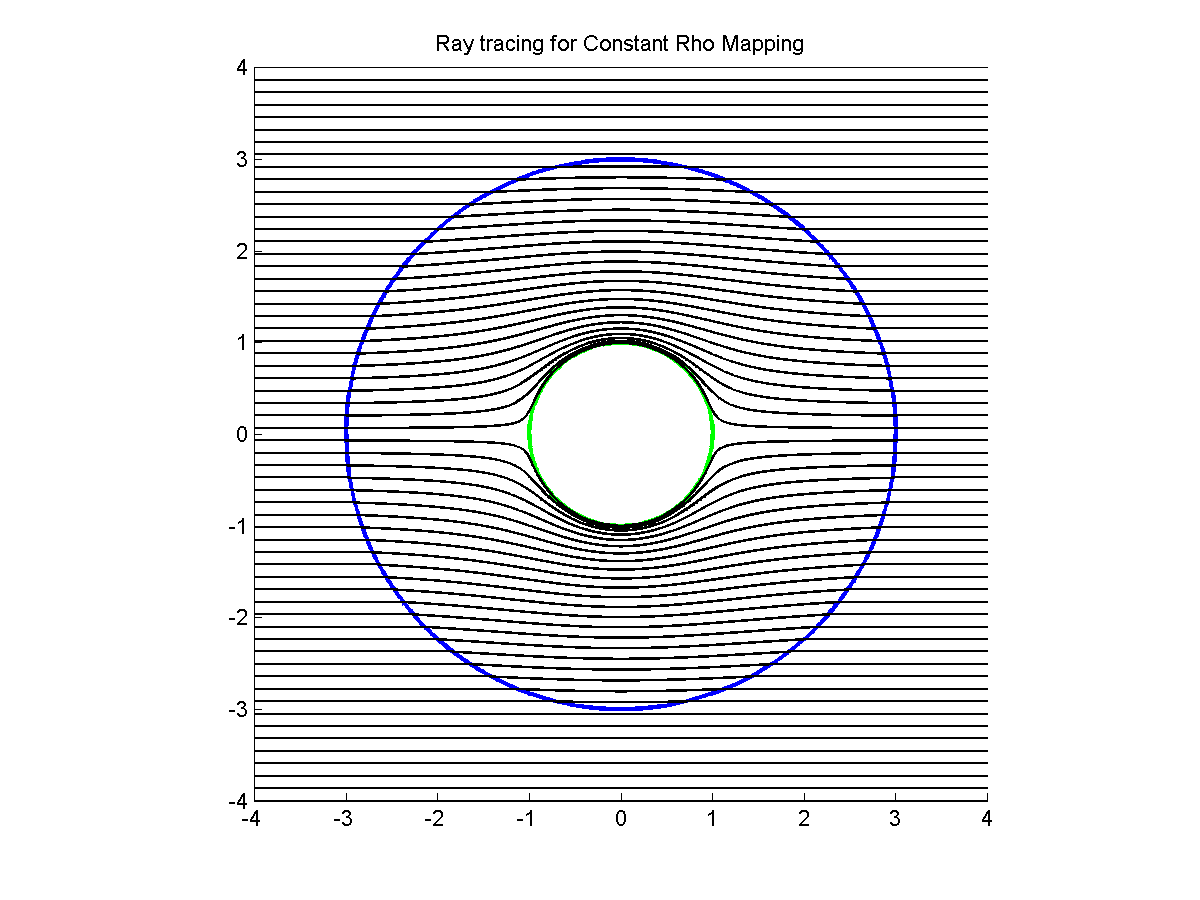}
}
\subfigure[\label{fig:raykrcons} Constant stiffness cloak]{
  \includegraphics[totalheight=\figtotalheight]{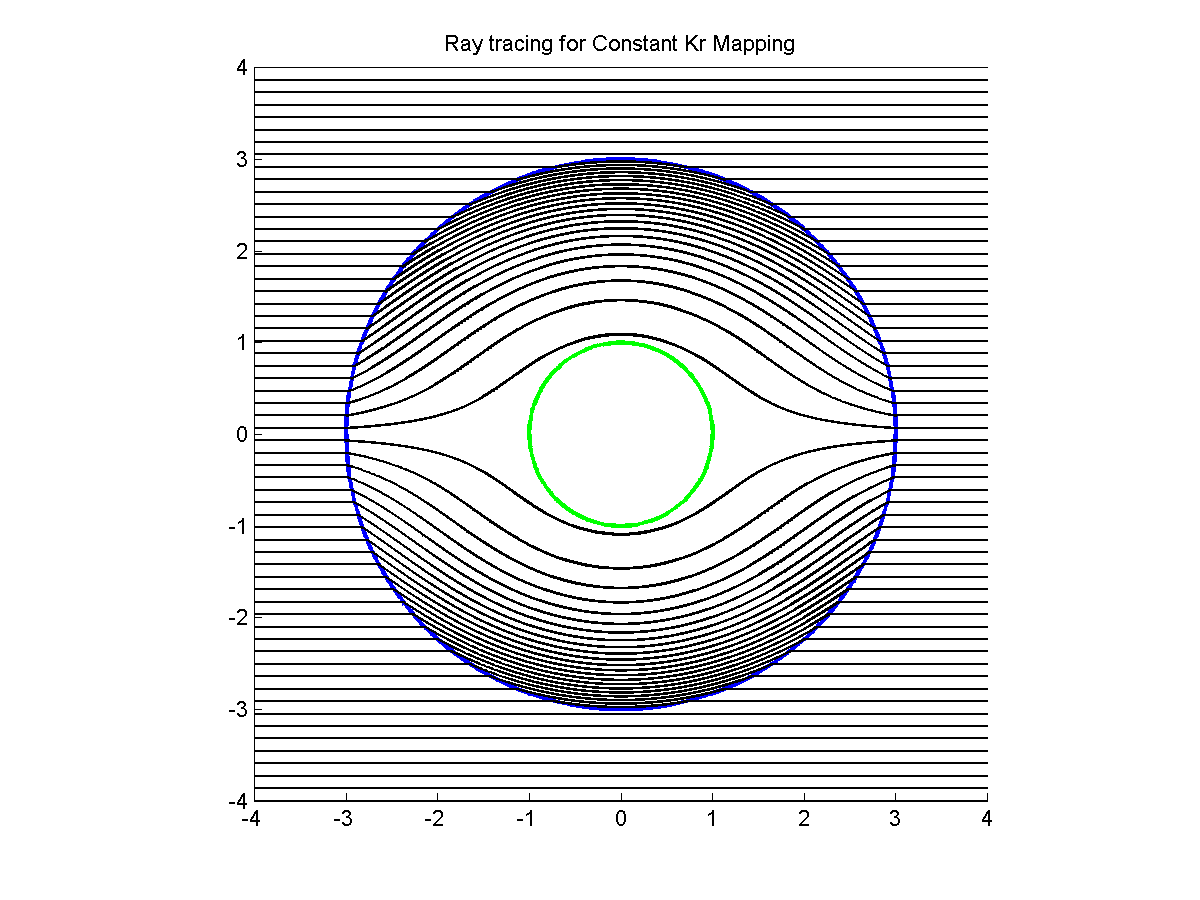}
}
\subfigure[\label{fig:rayoptimal} Cloak with optimal anisotropy]{
  \includegraphics[totalheight=\figtotalheight]{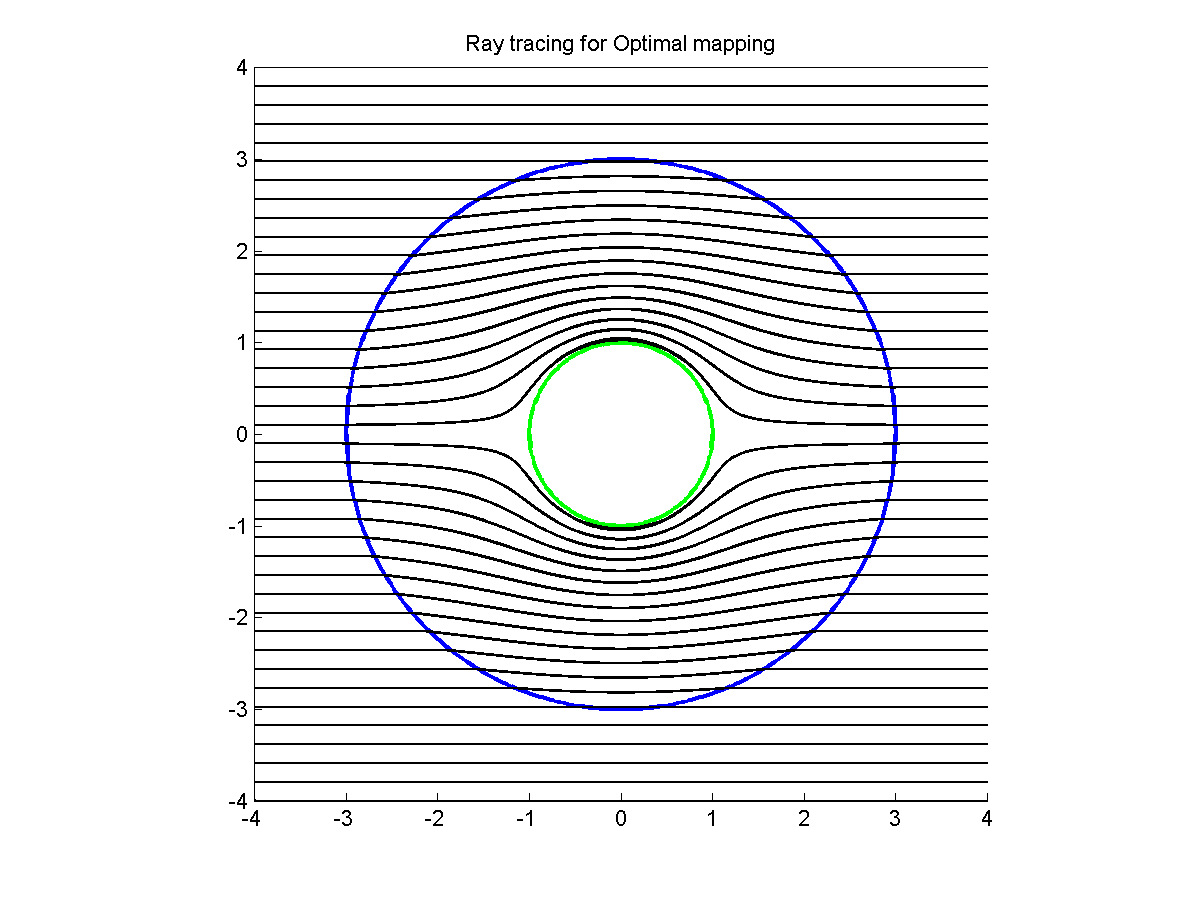}
}
\caption{\label{fig:rays1}Rays for KSVW, constant density, constant stiffness, and optimal anisotropy cloaks in $\text{2D}$, $a=1,b=3$.}
\end{figure}

We note that rays in the cloak are straight lines in deformed space\cite{Norris08b}. This allows us to trace the rays  for the constant density cloak by deforming the straight rays corresponding to a plane wave traveling through a homogeneous medium by the inverse transformation $f^{-1}$. Ray-tracing results are shown in Figure \ref{fig:rayrhocons}, from which we can see that the rays curve gently in the outer region of the cloak and sharply close to the inner radius. In our experience, the \textit{smooth} nature of this propagation makes it easy for this to be simulated with standard linear finite elements \footnote{Our finite element simulations were carried out with WAI's commercial finite element code PZFlex\textregistered, http://www.pzflex.com, last accessed \today}.
\begin{figure}
\centering
\subfigure[\label{fig:raypowerkr1} Initial propagation.]
{
  \includegraphics[totalheight=\figtotalheight]{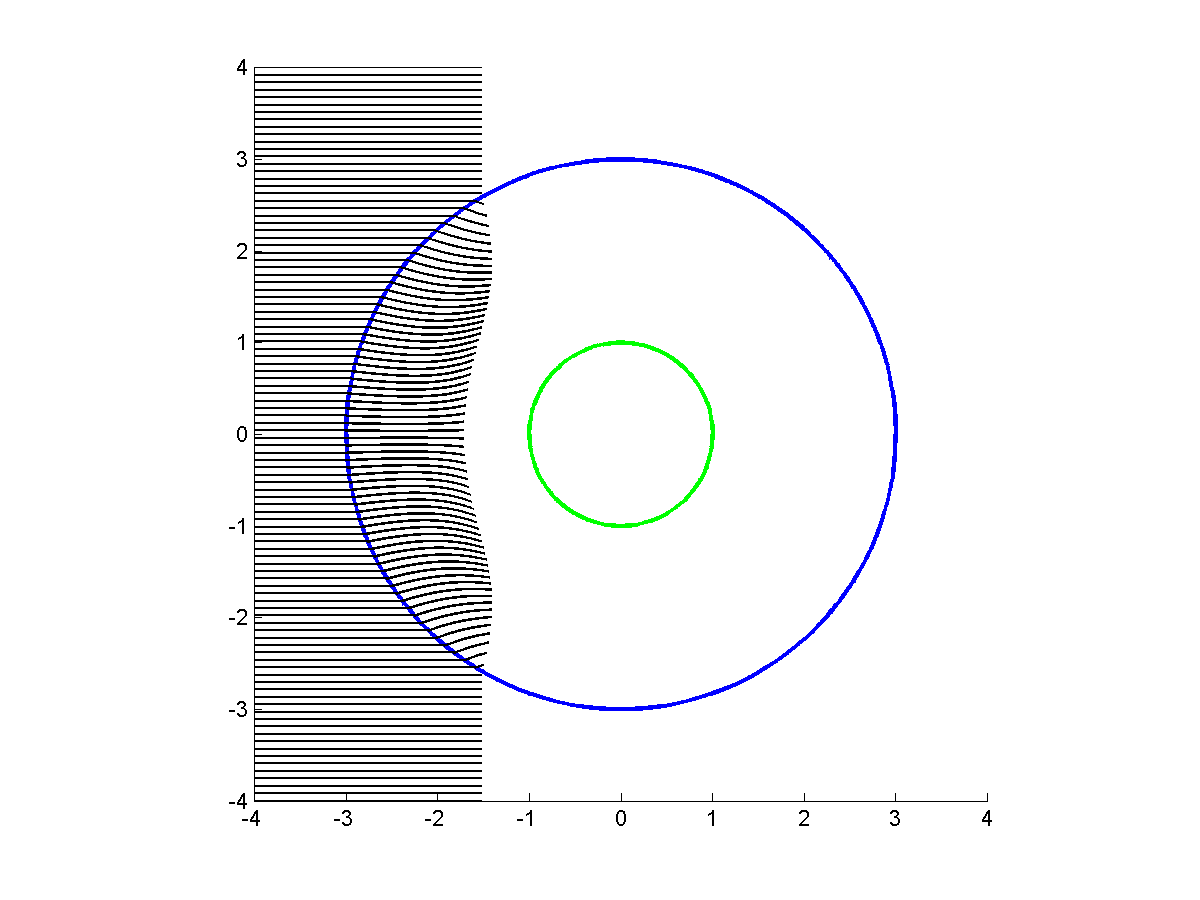}
}
\subfigure[\label{fig:raypowerkr2} Complete propagation.]
{
  \includegraphics[totalheight=\figtotalheight]{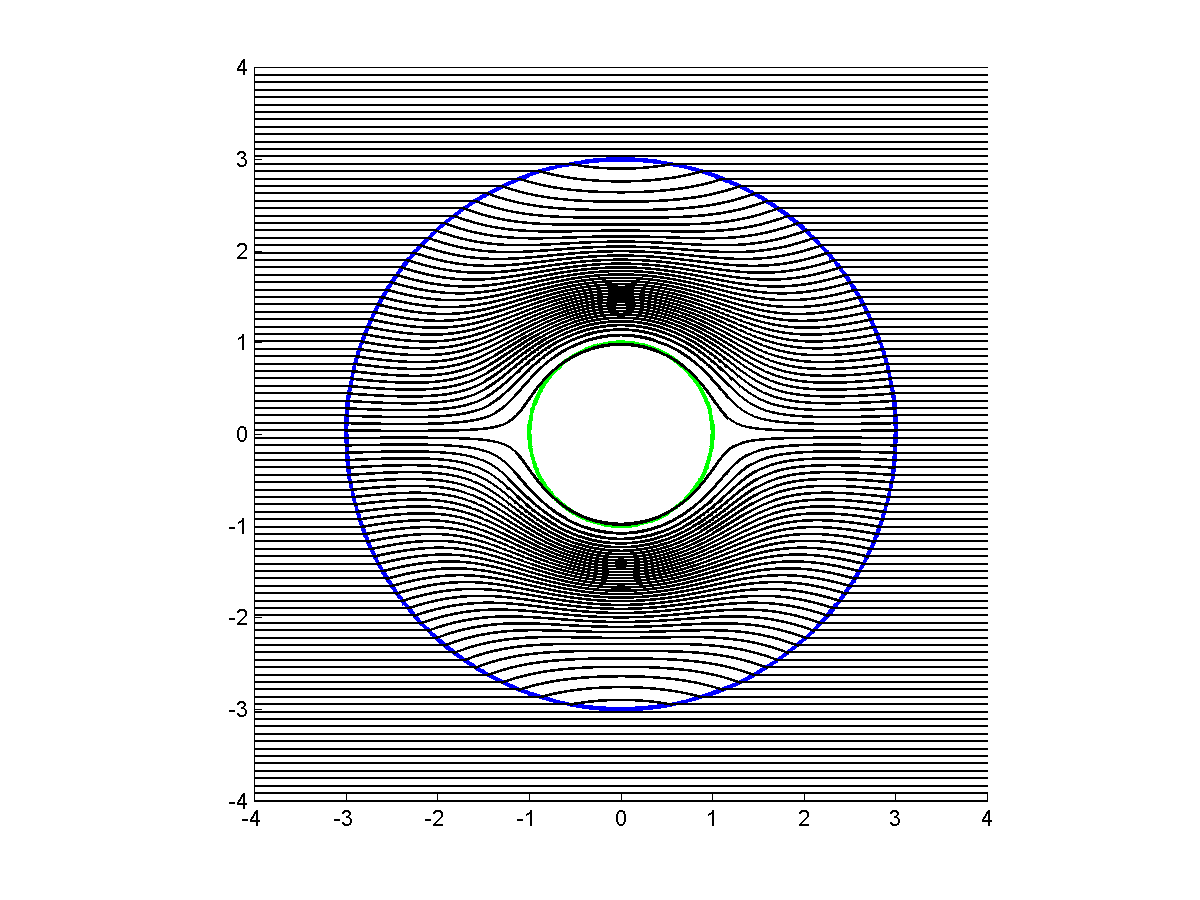}  
}
\caption{\label{rays:powerkr} Rays for power law $K_r$, with parameters $a=1,b=3$, $\alpha=3$.}
\end{figure}
\begin{figure}
\centering
\subfigure[\label{fig:raypowerrho1} Initial propagation.]
{
  \includegraphics[totalheight=\figtotalheight]{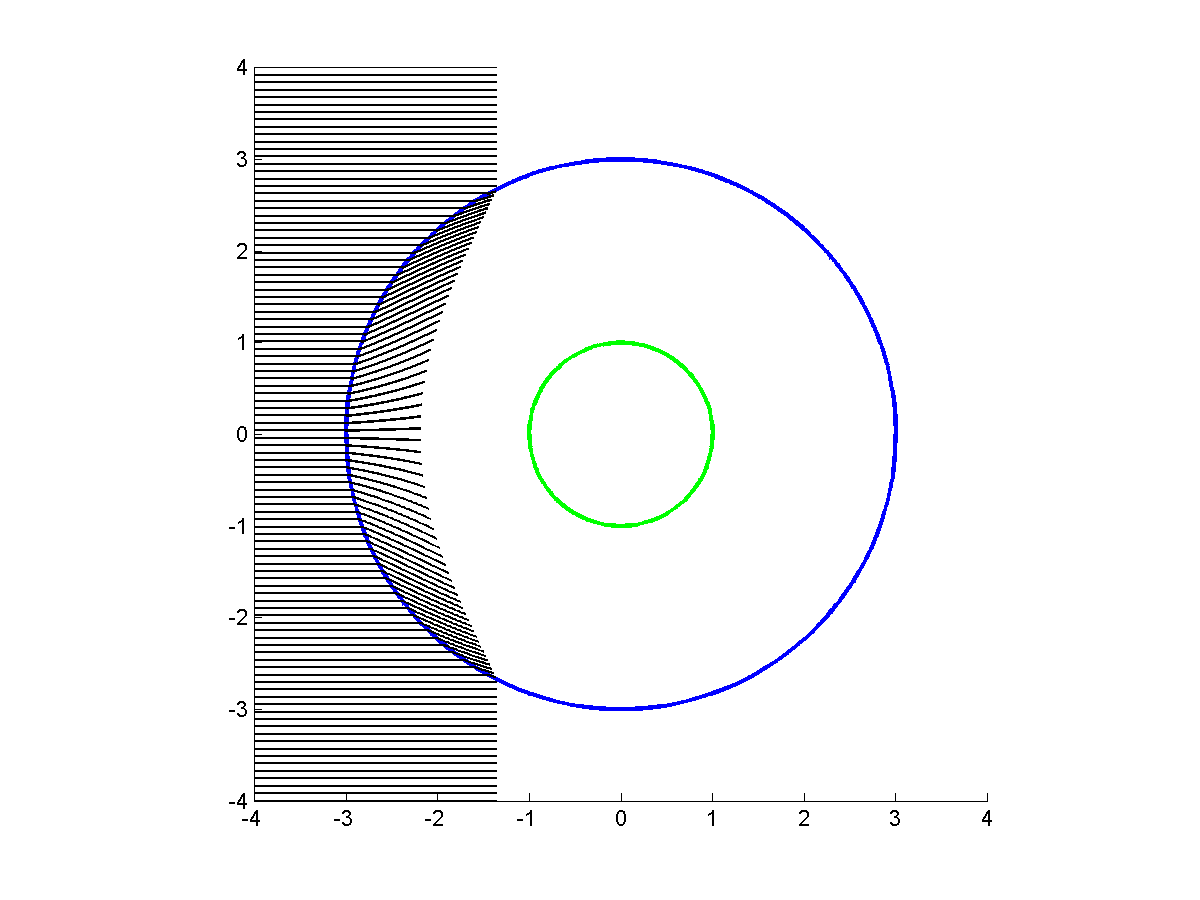}
}
\subfigure[\label{fig:raypowerrho2} Complete propagation.]
{
  \includegraphics[totalheight=\figtotalheight]{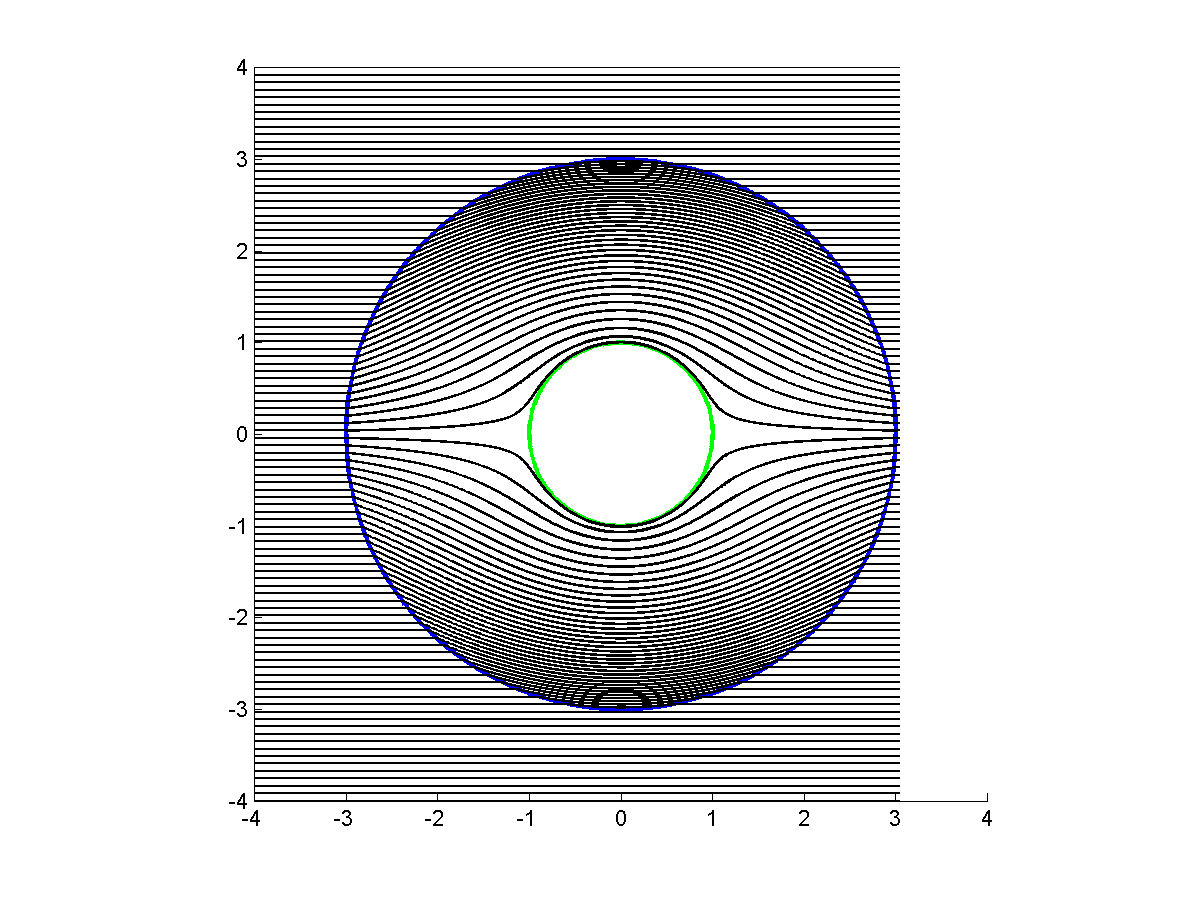}
}
\caption{\label{rays:powerrho1} Rays for power law $\rho$, with parameters  $a=1$, $b=3$, $\alpha=3$.}
\end{figure}
\begin{figure}
\centering
\subfigure[\label{fig:raypowerrho3} Initial propagation I.]
{
  \includegraphics[totalheight=\figtotalheight]{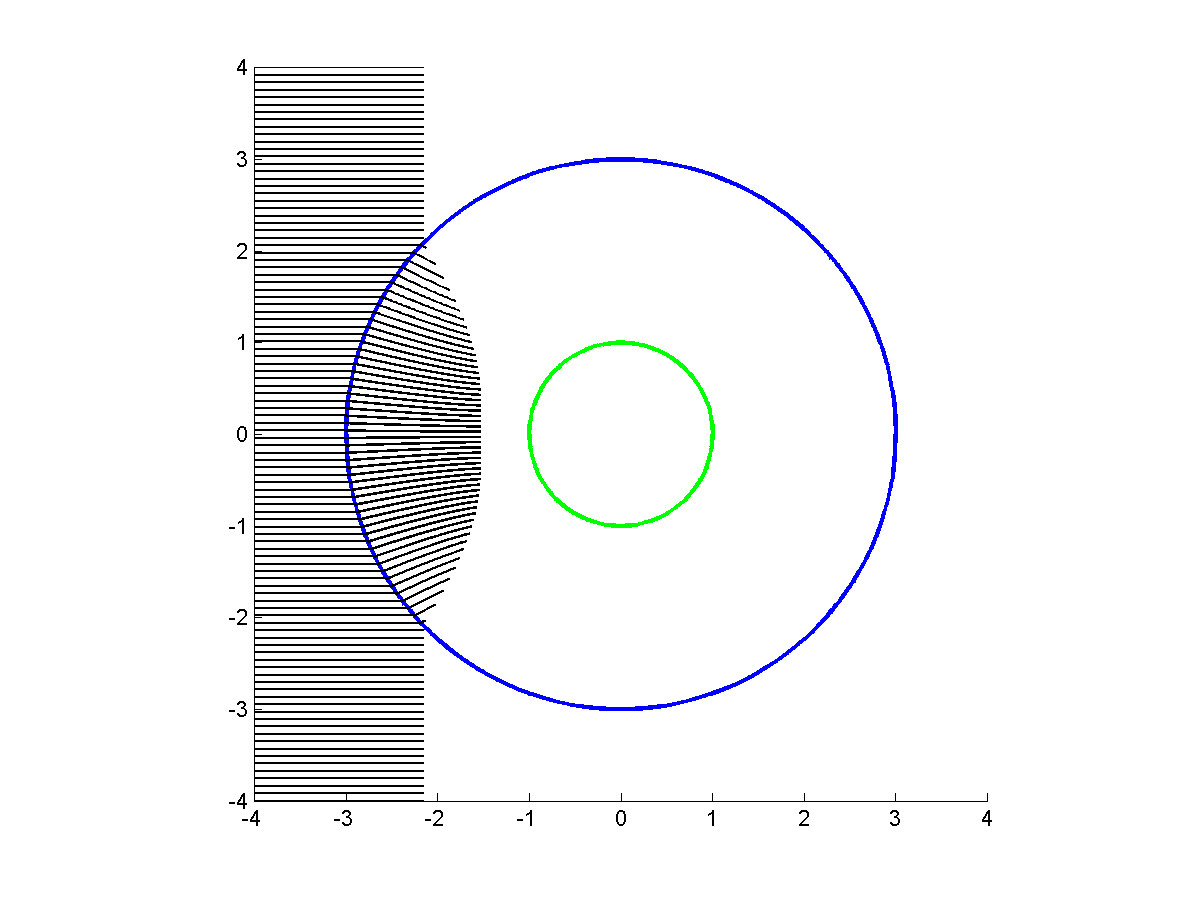}
}
\subfigure[\label{fig:raypowerrho4}  Initial propagation II.]
{
  \includegraphics[totalheight=\figtotalheight]{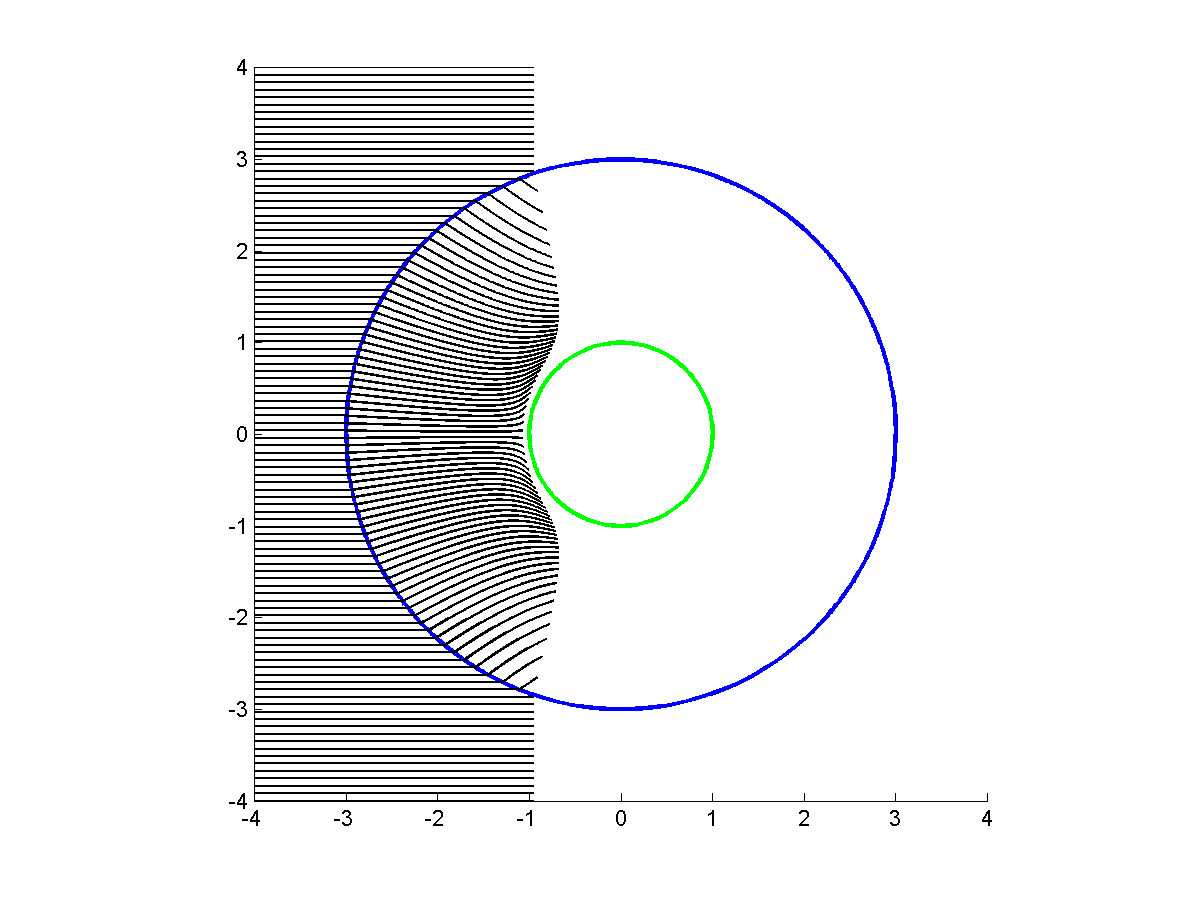}
}
\subfigure[\label{fig:raypowerrho6} Complete propagation.]
{
  \includegraphics[totalheight=\figtotalheight]{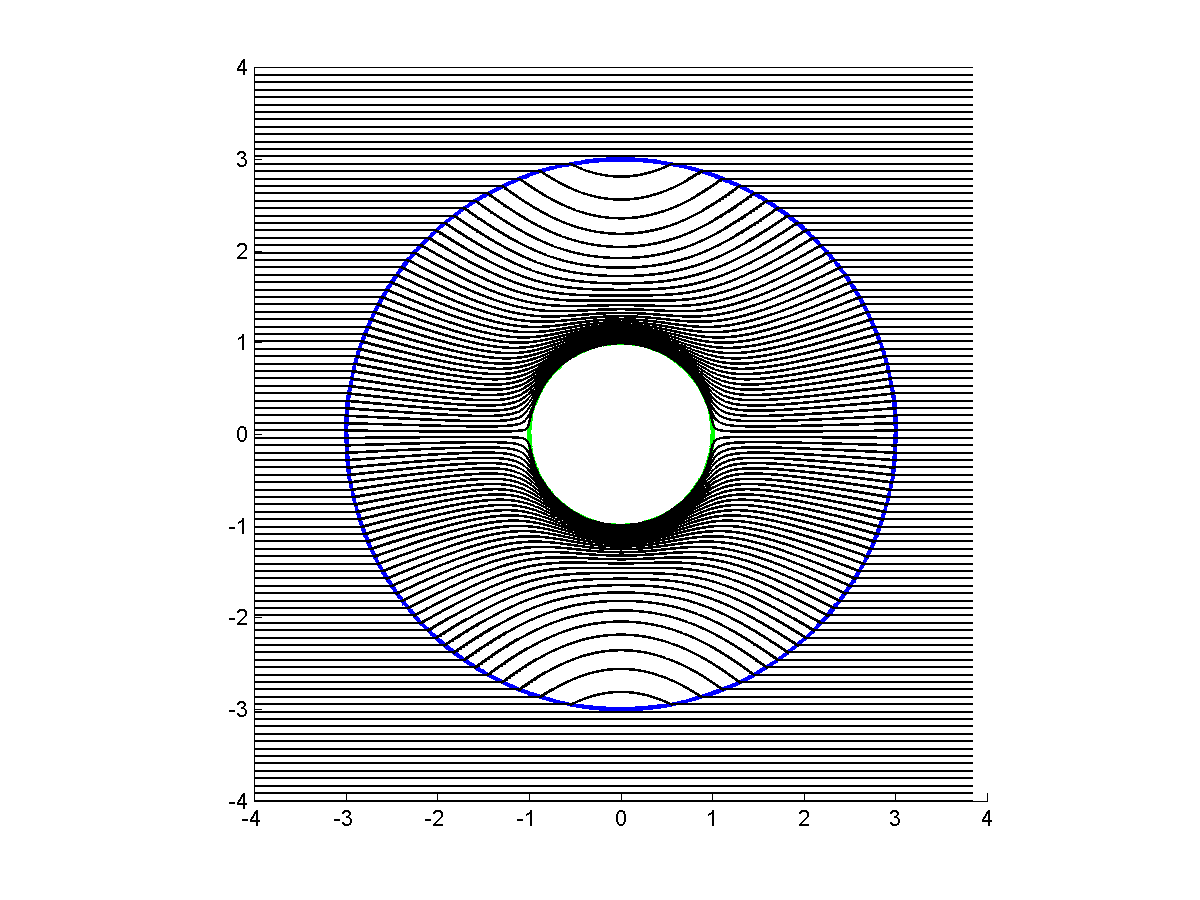}
}
\caption{\label{rays:powerrho2} Rays for power law $\rho$, with parameters  $a=1$, $b=3$, $\alpha=-3$.}
\end{figure}
\begin{figure}
\subfigure[\label{fig:raysuperlens1} Initial propagation. ]
{
  \includegraphics[totalheight=\figtotalheight]{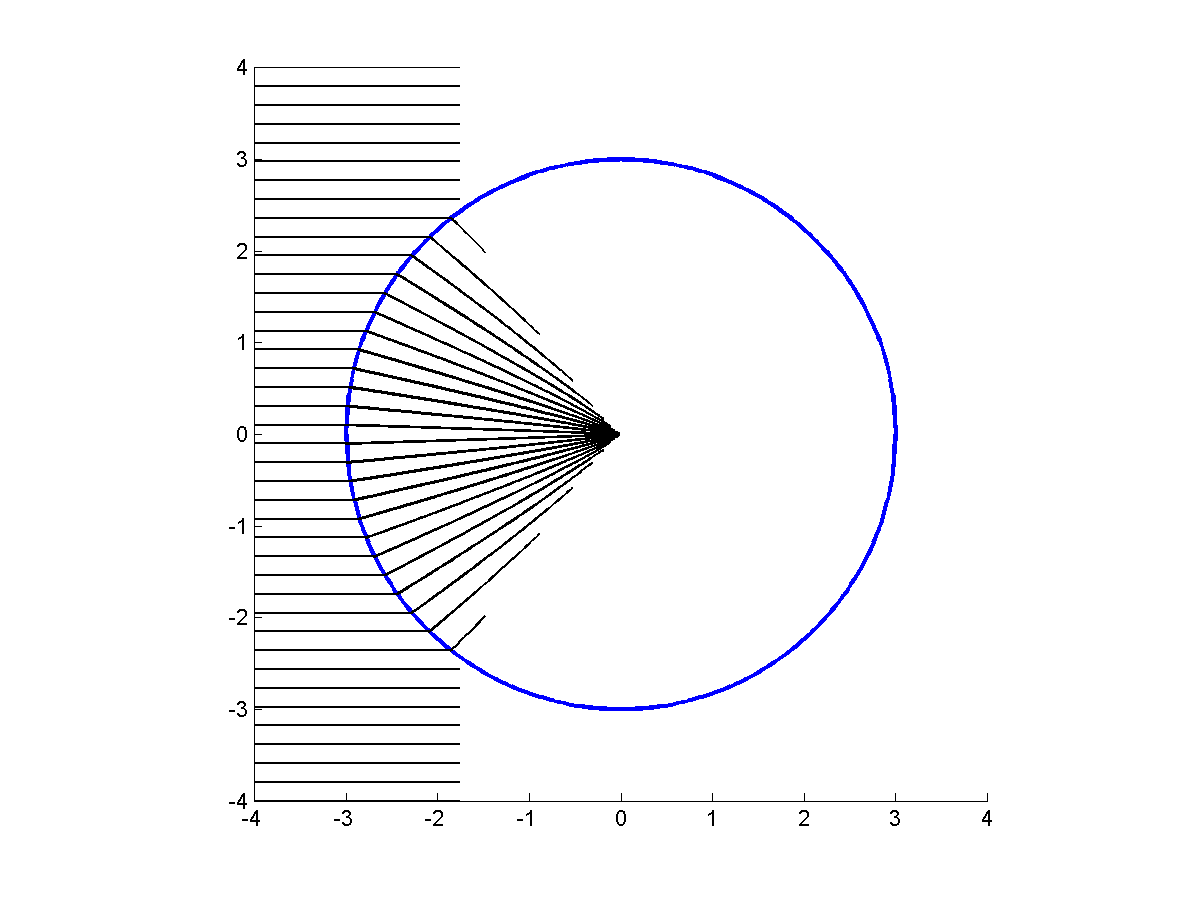}  
}
\subfigure[\label{fig:raysuperlens2} Rays converging. ]
{
  \includegraphics[totalheight=\figtotalheight]{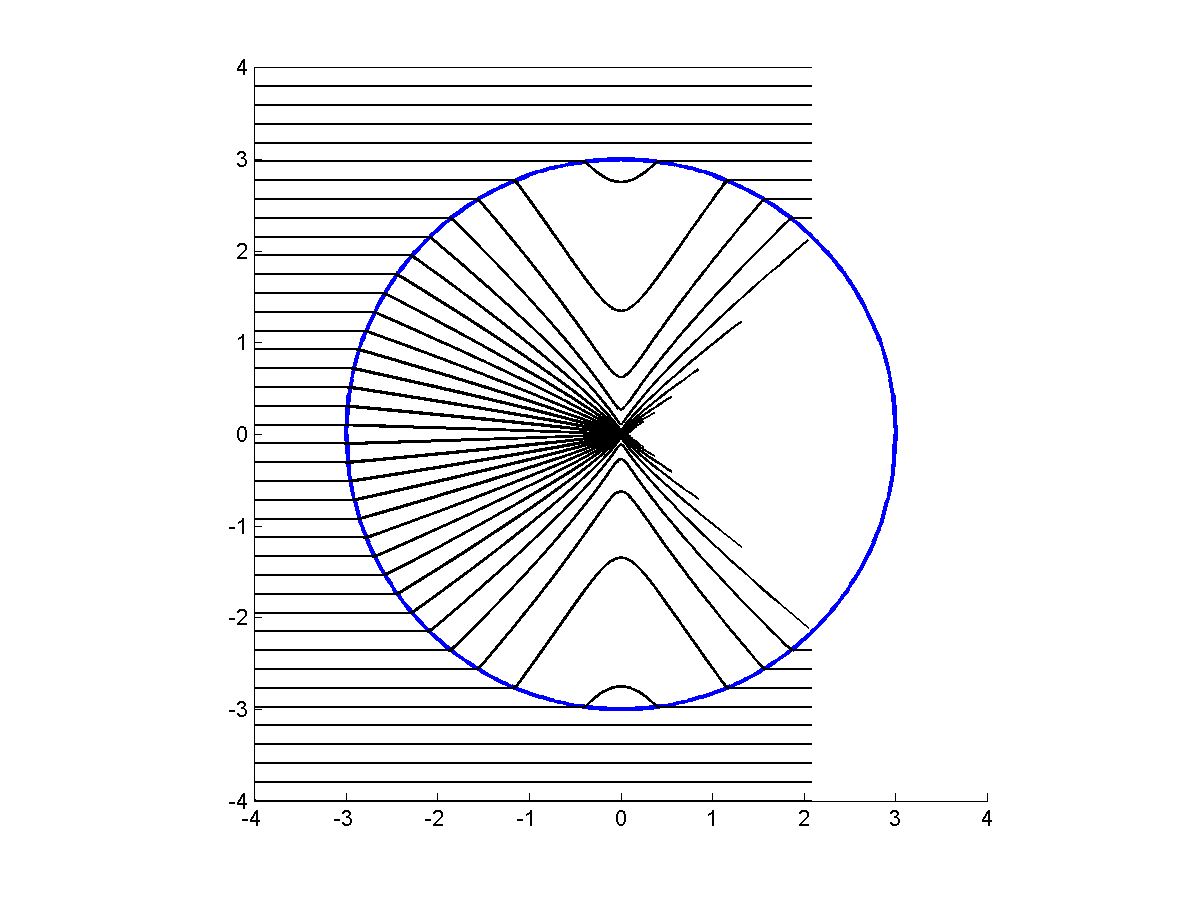}
}
\subfigure[\label{fig:raysuperlens3} Complete propagation. ]
{
  \includegraphics[totalheight=\figtotalheight]{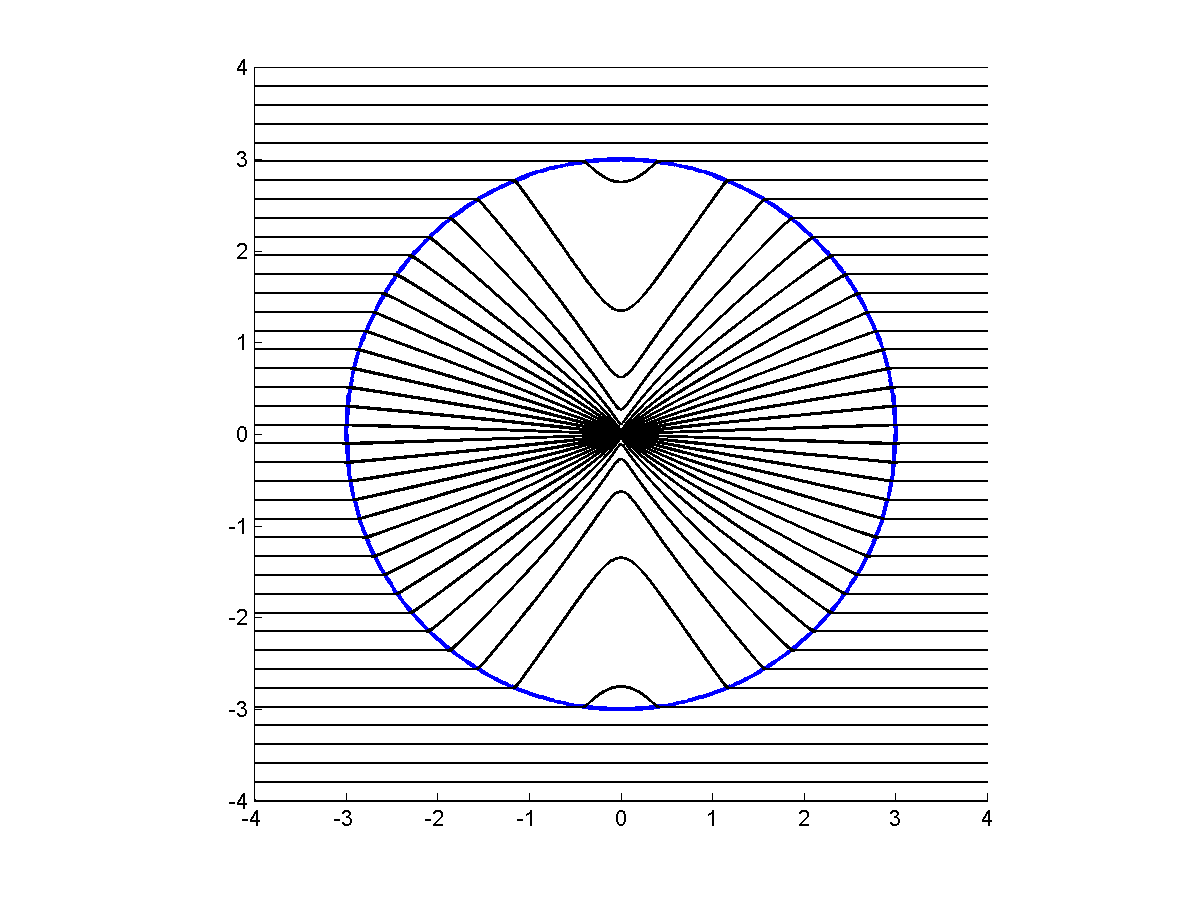}
}
\caption{\label{rays:superlens} Rays for an acoustic concentrator $f(r)=b(r/b)^{1/10}$.}
\end{figure}

\subsection{\label{sect:conskr}Transformations yielding constant radial stiffness $K_r$}
We formulate the problem by assuming that we are given $a > \delta > 0$ and $K_r$ constant. As before, start with the expression for $K_r$ from equation (\ref{eqn:pentak}), and treat it as an ordinary differential equation for $f$ and get 
\beq
\label{eqn:odeconskr}
\int{ \frac{f'}{f^{d-1}}dr} = {  \frac{K_0}{K_r}\int\frac{dr}{r^{d-1}} } .
\eeq
The solution is different for $\text{2D}$ and $\text{3D}$ and is determined separately in the next two subsections.

\subsubsection{\label{sect:2dconstkr}Transformation yielding constant $K_r$ in $\text{2D}$}
The transformation $f$ in $\text{2D}$ can be determined by solving equation (\ref{eqn:odeconskr}) to get 
\beq
\label{eqn:conskr2df}
f =  b\, \big(\frac{r}{b}\big)^{\frac{K_0}{K_r}} > 0, \ \ f'=  
\frac{K_0}{K_r} \frac fr > 0. 
\eeq
The cloak parameters then follow from eq.\  (\ref{eqn:pentak}) as
\beq\label{eqn:conskr2ddelta}
\begin{aligned}
&\rho    =  \rho_0 \frac{K_0}{K_r} \big(\frac{r}{b}\big)^{2(\frac{K_0}{K_r} -1)},\ \ 
K_r  = \frac{K^2_0}{K_\perp}= K_0\frac{\ln{a /b}}{\ln{\delta /b}} ,
\\
& \quad
\delta  =  a\, \big(\frac{a}{b}\big)^{\frac{K_0}{K_r} -1} ,\ \  \pdd{\delta}{K_r} = \frac{\delta}{K_r} \ln \big(\frac{b}{\delta}\big) >0.
\end{aligned}
\eeq
We prove that given $a> \delta >0$, it is possible to find $b$ such that $b>a>\delta>0 \, \forall\,\, K_r=\text{constant} > 0,\,K_r{\ne}K_0$. 
Since $\frac{\delta}{a}<1$ and $a<b$, the exponent $\frac{K_0}{K_r} -1$ in eq.\ (\ref{eqn:conskr2ddelta}) must be positive to ensure $b>a$. i.e. $0< K_r < K_0$ must hold for constant $K_r$ cloaks in $\text{2D}$. This shows that the cloak outer radius is greater than the inner radius, making the cloak physically realistic.

This case is of interest because the only parameter that varies with $r$ is the density.  At the outer radius $\rho  (b)  =  \rho_0 \frac{K_0}{K_r} >  \rho_0 $, and the value at the inner radius is $\rho  (a)  =  \rho  (b) (\delta /a)^2$.

The rays for a constant $K_r$ cloak in $\text{2D}$ are shown in Figure \ref{fig:raykrcons}. Unlike the rays for the constant density cloak shown in Figure \ref{fig:rayrhocons}, the rays for constant stiffness curve sharply at the outer surface of the cloak.  In our experience, the propagation of waves near the surface is extremely hard to capture with standard linear, time-domain, finite elements, possibly due to sharp change in the direction of propagation. The required element density is in the order of hundreds of elements per shortest wavelength. This is contrast to the rule of thumb in transient, explicit finite element analysis in which typically sixteen elements per shortest wavelength are used.
\subsubsection{\label{sect:3dconstkr}Transformation yielding constant $K_r$ in $\text{3D}$}
Treating equation (\ref{eqn:odeconskr}) with $d=3$ as a differential equation for $f$ with $K_r$ known and constant, gives
\beq
f= \big[ \frac 1b +\frac{K_0}{K_r}\Big(\frac{1}{r} -\frac{1}{b}\Big) \Big]^{-1}  
\ \text{ and } \ f'=\frac{K_0}{K_r}\big(\frac{f}{r}\big)^{2} > 0.
\eeq
Using the above definition of $f$ in eq.\ (\ref{eqn:pentak})  we get
\bse
\ber
\rho &=& 
{ \rho_0\frac{K_0}{K_r}\bigg(\frac{bK_r}{bK_0 + r(K_r-K_0)}\bigg)^4 }, 
\\
K_\perp &=&  
{ K_r \bigg(\frac{bK_0}{bK_0 + r(K_r-K_0)}\bigg)^2}, 
 \\
\label{eqn:conskr3ddelta}
\delta &=&  f(a) = \frac{abK_r}{bK_0 + a(K_r-K_0)} .
\eer
\ese
We now prove that under certain conditions it is possible to find $b$ such that $b>a>\delta>0$, meaning that the cloak outer radius is greater than the cloak inner radius. As a consequence we prove that $f>0$. We start by rewriting equation (\ref{eqn:conskr3ddelta}) as 
\beq
\frac{b}{a} =   \big(  \frac{K_0}{K_r}-1 \big) /\big(\frac{K_0}{K_r} - \frac{a}{\delta}\big) .
\eeq
To ensure $\frac{b}{a}>1$, we require $K_r < K_0\frac{\delta}{a}$. Since, $f'>0 \in (a,b)$ and $f(a)=\delta>0$, it follows that $f$ is monotonic in the interval $(a,b)$ and therefore $f>0 \,\in\,(a,b)$. Note that, if $\delta << a \, \implies K_r << K_0$ . Since $K_\perp = K_0^2/K_r$, this means $K_\perp >> K_r$. Thus, to ensure a small scattering cross section we need a material which is very stiff in one direction ($\perp$) as compared to the other ($r$). We expect that such a material will require careful engineering, and such practical constraints may restrict the amount of reduction in the scattering cross section that can be achieved.

\subsection{\label{sect:conskt}Transformations with constant tangential stiffness $K_\perp$}
\subsubsection{Transformations yielding constant $K_\perp$ in $\text{2D}$}
In two dimensions, constant $K_\perp$ is the same as constant $K_r$ (which was previously considered), because $K_rK_\perp=K_0^2$ in $\text{2D}$.
\subsubsection{Transformations yielding constant tangential stiffness in $\text{3D}$}
We formulate the problem as follows. We consider that we are given $a>\delta>0$. We need to find a $f>0,f'>0$ satisfying $f(a)=\delta$, $f(b)=b$, $b>a$. We treat eq.\ (\ref{eqn:pentak}) as a differential equation for $f$ and obtain after using the boundary condition $f(a)=\delta$,
\beq\label{1=2}
f(r) = \delta + \frac{K_\perp}{K_0}(r-a) > 0 \ \  
\text{ and }f' = \frac{K_\perp}{K_0} > 0, 
\eeq
for $r \ge a$. 
Using $b= f(b)  >a$ yields the condition $K_\perp>K_0$. The constant $K_\perp$ cloak in three dimensions is characterized by $f$ of eq.\ \eqref{1=2} and 
\beq
\begin{aligned}
\rho &= \rho_{0}\frac{K_\perp}{K_0}\Big(\frac{\delta}{r} + \frac{K_\perp}{rK_0}(r-a) \Big)^{2},
\\ 
K_r &= \frac{K^3_{0}}{K_\perp^2} \frac{\rho}{\rho_0}, \ \ 
\delta = b - \frac{K_\perp}{K_0}(b-a) .
\end{aligned}
\eeq
Note that this is the same transformation as the KSVW transformation in eq.\ (\ref{eqn:threetransformations}a).

\section{\label{sect:powerlaw}Transformations yielding power law property variation in $\text{2D}$}

We now consider more complicated spatial distributions for the material properties, namely power law variations of density, radial stiffness tangential stiffness and a case in which density is proportional to stiffness. This treatment is restricted to two dimensions. 
\subsection{Transformations yielding power law density}
Consider the following strategy: Given $b>a>\delta >0$, $\alpha \ne 0$, we determine $\rho(a)=\rho_a$ that ensures  $f(b)=b$ and prove that $\rho_a >0$. The power law for density we consider is as follows:
\beq
\label{eqn:rhopower}
\rho(r) 
= \rho_{a}\Big(\frac{r}{a}\Big)^{\alpha} \ \Rightarrow \ 
\int{ff'dr} =  \frac{\rho_a}{\rho_0}\int{\frac{r^{\alpha + 1}}{a^{\alpha}} dr}, 
\eeq
where the latter is a consequence of eq.\ (\ref{eqn:pentak}). 
The cases $\alpha\ne-2$ and $\alpha=-2$ need to be considered separately. 
To summarize, the power law density cloak in two dimensions is characterized by \eqref{eqn:rhopower}$_1$, with 
\beq
\begin{aligned}
f^2&=\delta^{2} +  2\frac{\rho_a}{\rho_0}\frac{(r^{\alpha+2}-a^{\alpha+2})}{a^{\alpha}(\alpha+2)} ,
\\
\rho_a&= \rho_0 \, \frac{(\alpha+2)}2 \, \frac{a^{\alpha}(b^{2} - \delta^{2})}{(b^{\alpha+2} - a^{\alpha+2})},
\end{aligned}
\ \ \text{for }
\alpha \ne-2,
\eeq
and 
\beq
\begin{aligned}
f^2 &= \delta^{2} +2 a^2\frac{\rho_a}{\rho_0}\ln(\frac{r}{a}) ,
\\
\rho_a &= \rho_0\frac{(b^2 - \delta^2)}{2a^2 \ln{\frac{b}{a}}},
\end{aligned}
\ \ \text{for }
\alpha=-2,
\eeq
and in both cases, 
\beq
f' =  \frac{r^{\alpha +1} \rho_a }{f a^{\alpha} \rho_0},
\ \ 
K_r = K_0 \, \frac{\rho_0}{\rho_a} \, \frac{a^{\alpha} f^2}{ r^{\alpha+2} }.
\eeq
Note that the $\alpha =0$ corresponds to the special case of constant density in $\text{2D}$, considered in 
\S\ref{sect:consrho}.  
To prove $\rho_a>0$ for $\alpha \ne -2 $, consider the two cases $\alpha > -2 $ and $\alpha < -2$. In the first case we have, $b^{\alpha+2} > a^{\alpha+2}$, and therefore $\rho_a > 0$. In the second case, we have, $b^{\alpha+2} < a^{\alpha+2}$. 
In addition, $\alpha+2<0$ and therefore 
$
(\alpha+2)/( b^{\alpha+2} - a^{\alpha+2}) > 0$. 
Hence $\rho_a > 0$ in this case as well. The same argument, substituting $r$ for $b$ can be used to prove $f^2>0$, and choosing the positive root, we get $f>0$. The positivity of $f'$ follows.  
A similar analysis can be performed for $\alpha=-2$. 
\subsection{Transformations yielding power law radial stiffness $K_r$}
Here, we consider the following power law for $K_r$
\beq
\label{eqn:powerkr}
K_r(r) = K_a\Big(\frac{r}{a}\Big)^{\alpha}, \quad \alpha \ne 0 .
\eeq
The case $\alpha = 0$ (constant $K_r$) was considered in \S\ref{sect:conskr}. 
Given $b > a > \delta$, $\alpha\ne{0}$, we find what value of $K_a$ ensures that $f(b)=b$. We then prove this $K_a>0$. 
Using eqs.\ (\ref{eqn:pentak}), (\ref{eqn:powerkr})  plus the boundary conditions $f(a)=\delta$ and $f(b)=b$, we get 
\bse
\ber
\label{eqn:powerkrf}
f &=& \delta\, \exp \big\{\frac{K_0}{K_a\alpha}\Big(1 - \Big(\frac{a}{r}\Big)^{\alpha}  \Big) \big\}  ,\\
f'&=& f\frac{K_0}{K_a}\frac{a^{\alpha}}{r^{\alpha+1} }  ,
\\
K_a &=& \frac{K_0}{\alpha\ln(\frac{b}{\delta})}\Big(1 - \Big(\frac{a}{b}\Big)^{\alpha}\Big) ,
\\
\rho &=& \rho_0\frac{f^2}{r}\frac{K_0}{K_a} \frac{a^{\alpha}}{r^{\alpha+1} } ,
\eer
\ese
each of which is clearly positive.

\subsection{An acoustic concentrator}
Motivated by the form of equation (\ref{eqn:powerkrf}), we consider the transformation $r=b(R/b)^{10}$ or equivalently, $R=f(r)=b(r/b)^{1/10}$, yielding the rays shown in Figure \ref{rays:superlens}. The focus of the rays can be made arbitrarily tight. The process can  also be reversed: one can place a source at the focus and convert a cylindrical wavefront generated by a point source into a plane wavefront. Similar work on designing acoustic concentrators using transformation acoustics has been recently reported in Wang \textit{et al.}\cite{Wang:concentrator}. Previously, Rahm \textit{et al.}\cite{Rahm07} reported the design of electromagnetic concentrators. 
\subsection{Transformations yielding power law tangential stiffness}
In two dimensions, this is equivalent to power law $K_r$, because $K_\perp = {K_0^2}/{K_r}$.

\subsection{Transformations yielding proportional density and radial stiffness}
Since density is usually associated with stiffness, we consider a power law linking density with stiffness in the radial direction. This power law is defined as 
\beq \label{2=2}
K_r = \beta\rho^{\alpha},
\eeq
for $\alpha$, $\beta (>0)$ constant. 
Using the pentamode relations (\ref{eqn:pentak})  for $K_r$ and $\rho$ in the above equation, we get a differential equation for $f$
\beq\label{123}
f^{\lambda -1} f'= \omega^{\lambda}\, r^{\lambda -1}
\eeq
where
\beq
\lambda = \frac{\alpha d +2 - d}{\alpha+1}, 
\quad 
 \omega =  \Big( \frac{K_0}{\beta\rho_{0}^{\alpha}}\Big)^{\frac{1}{(\alpha+1)\lambda} }.
\eeq
When $\lambda \ne  0$ eq.\ \eqref{123} yields
\beq
f(r) = \Big(\delta^{\lambda} + \omega^{\lambda} \big(r^\lambda - a^\lambda)\Big)^{1/\lambda}
\eeq
This equation is not consistent with the constraint that $f(b)=b >a$ if $\lambda <0$, in general.  For $\lambda >0$, setting $f(b) = b$ yields    
\beq
\delta = \Big(b^{\lambda}- \omega^{\lambda} \big( b^\lambda - a^\lambda \big)\Big)^{\lambda}
\eeq
Requiring that $0< \delta <a$ implies the a range of possible values for $\beta$.
The special case of $\lambda = 0$, i.e. $\alpha = (d-2)/3$,  needs to be distinguished. 
In $\text{2D}$, this means $\alpha=0$ and therefore $K_r=constant$, and is therefore not interesting. In $\text{3D}$, this implies, $\alpha=\frac{1}{3}$, and eq.\ \eqref{123} can be integrated to give a power law solution for $f(r)$, 
\beq
f= b\, \big( \frac rb\big)^\mu, 
\ \ 
\mu = \Big( \frac{K_0}{\beta\rho_{0}^{1/3}}\Big)^{3/4} 
\quad (d=3,\ \alpha = \frac 13),
\eeq
In this case, $\delta = b\, \big( \frac ab\big)^\mu $, which is clearly positive, satisfies the constraint
  $\delta < a$ only if $\mu >1$, or equivalently, $\beta  < K_0 /\rho_{0}^{1/3}$, setting an upper limit on 
$\beta$. 

In summary, eq.\ \eqref{2=2} has cloak-like solutions for $\alpha \ge 0$ in $\text{2D}$, and $\alpha \ge \frac 13$ in $\text{3D}$, with associated limits on the possible range in value of the parameter $\beta$.

\section{\label{sect:optimalaniso}Transformations yielding minimal elastic anisotropy }
Here, we consider acoustic cloaks which have minimal elastic anisotropy in a certain sense. We are motivated by the fact that extremely anisotropic materials are hard to design and manufacture. Minimizing anisotropy therefore may lead to a practical cloak. 
We begin by  defining two measures of anisotropy in equation (\ref{eqn:twomeasures}) and prove that only one of them yields physically meaningful transformations. 

\subsection{Optimal transformations in cylindrical  cloaks}

We define the following parameter
to be a local measure of the anisotropy in the cloak,  
\beq
\label{eqn:alphadef}
\gamma =\alpha + \frac 1\alpha, \quad \text{ where }
\alpha = \frac{c_{r}}{c_\perp}  
\eeq
This is the same  anisotropy parameter introduced in \cite{Pendry08}.  It can be shown that the minimum value of eq.\ (\ref{eqn:alphadef})  is $2$ and it occurs for  $\alpha=1$, i.e. when there is no anisotropy.
Based on eq.\ (\ref{eqn:alphadef}) we introduce two global measures of   cloak anisotropy, 
\beq
\label{eqn:twomeasures}
\gamma_1 =  V_\omega^{-1}\int_{\Omega} \gamma \, \dd V,
\qquad 
\gamma_2 =  V_\Omega^{-1}\int_{\omega} \gamma \, \dd v,
\eeq
where $V_\omega$, $V_\Omega$, are the volumes (areas) in the physical and virtual domains, respectively. 
It follows from eq.\ \eqref{1=1} and the identity 
\eqref{3=3} for d=2 $( K_0^2 = K_r K_\perp )$ that
\beq\label{4=4}
 J^{-1}\tr{\bd{B}} = \frac{K_r}{K_0} + \frac{K_\perp}{K_0} 
= \sqrt{\frac{K_r}{K_\perp}} + \sqrt{\frac{K_\perp}{K_r}} =\gamma  .
\eeq
Similarly, 
\beq\label{5}
 J\tr{\bd{B}^{-1}} = \frac{K_0}{K_r} + \frac{K_0} {K_\perp}
= \sqrt{\frac{K_r}{K_\perp}} + \sqrt{\frac{K_\perp}{K_r}} =\gamma  .
\eeq
Based on the identities \eqref{4=4} and \eqref{5}, it follows that  
\bse\label{5=5}
\ber
\label{eqn:gamma1}
\gamma_1 & =  V_\omega^{-1}\int_{\omega} \tr {\bd B}^{-1}  \, \dd v
=  V_\omega^{-1} \int_{\omega}   \frac{\partial X_i}{\partial x_j}\frac{\partial X_i}{\partial x_j} \, \dd v,
\\
\label{eqn:gamma2}
\gamma_2 & =  V_\Omega^{-1}\int_{\Omega}  \tr {\bd B}\, \dd V 
=  V_\Omega^{-1}\int_{\Omega}  \frac{\partial x_i}{\partial X_j}\frac{\partial x_i}{\partial X_j}\, \dd V .
\eer
\ese
The parameter $\gamma_1$ is therefore  the average in the current configuration of the sum of the principal stretches of the mapping from the original (virtual) domain.  Conversely, $\gamma_2$ is  the average in the original configuration of the sum of the principal stretches of the inverse mapping from the current (spatial)  domain. 

The global anisotropy measures $\gamma_1$ and $\gamma_2$ are minimized by the Euler-Lagrange equations.  
Consider $\gamma_1 $, then assuming $\omega$ is fixed, we have 
\ber
\delta \gamma_1   &=& 2
V_\omega^{-1} \int_{\omega}   \frac{\partial X_i}{\partial x_j}\delta \big( \frac{\partial X_i}{\partial x_j} \big)\, \dd v 
\nonumber \\ \nonumber
&=& \frac{2}{
V_\omega } \Big{(}\int_{\partial \omega}   \frac{\partial X_i}{\partial x_j}\delta   {  X_i}\, n_j \dd s -  
\int_{ \omega}   \frac{\partial^2 X_i}{\partial x_j\partial x_j }\delta   {  X_i}  \, \Big{)}
\dd v . 
\eer
The surface integral vanishes because, by assumption, the value of ${\bd X}$ on the boundary of $V_\omega$ is constant (in fact ${\bd X} = {\bd x}$ is required on $\partial \omega$), and therefore we deduce
\bse
\ber
\label{eqn:lapcapX}
\text{min}\, \gamma_1 & \quad \Leftrightarrow \quad \nabla^2 {\bd X} = 0 \, \text{ in }\omega,\\
\label{eqn:lapsmallx}
\text{min}\, \gamma_2 & \quad \Leftrightarrow \quad \nabla_{X}^2 {\bd x} = 0 \, \text{ in }\Omega .
\eer
\ese
It is interesting to note that these equations are satisfied by conformal transformations, a large class of potential transformations.  Here, however, we restrict attention to purely radial transformations. 

\begin{figure}[tb]   
\begin{center} 
\includegraphics[totalheight=\figtotalheight]{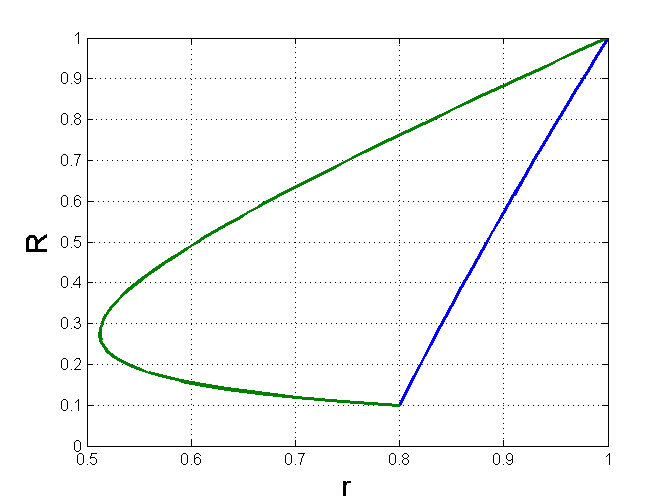}
\end{center} 
\caption{\label{fig:fig1}The solid curve shows the transformation defined by  (\ref{eqn:defR2d}) for $\{\delta, a,b\}=\{0.1,0.8,1.0\}$.  The dashed curve is the mapping (\ref{eqn:defr2d}) for the same cloak parameters.}
\end{figure}

Consider \eqref{eqn:lapcapX} first.  Assuming the inverse mapping ${\bd X} = f(r) r^{-1}{\bd x}$ then it is straightforward to show that 
$\nabla^2 {\bd X} = [ r(rf')'-f] (fr^2)^{-1} {\bd X}$, and (\ref{eqn:lapcapX}) is satisfied if $f = Ar+Br^{-1}$, for constants $A$ and $B$.  As before, we assume the cloak occupies $R\in [\delta , b]$,  $r\in [a , b]$
with $0 < \delta < a < b$. 
The constants are then found from the conditions $f(a)=\delta$ and $f(b)=b$, yielding 
\beq
\label{eqn:defR2d}
f(r)=(b^2-a^2)^{-1}\bigg[ (b^2-a\delta)r-(a-\delta)b^2 \frac{a}{r} 
\bigg].
\eeq
The same result can be found by noting that the anisotropy parameter of  (\ref{eqn:gamma1}) reduces for radially symmetric transformations to 
\beq
\gamma_1 = 
\frac {d}{b^d-a^d} \int_{a}^{b}\bigg[
(f')^2 + (d-1)\bigg( \frac f r \bigg)^2\bigg] r^{d-1} \dd r,  
\eeq
for $d=2$ or $d=3$. 
The minimizer satisfies the Euler-Lagrange equation $r^{3-d}(r^{d-1}f')'-(d-1)f=0$, which for $d=2$ gives (\ref{eqn:defR2d}). In the same way, we find that (\ref{eqn:lapsmallx}) is satisfied if $r = AR+BR^{-1}$, for constants $A$ and $B$. The end  conditions $r(\delta )=a$ and $r(b)=b$ imply that the transformation which  minimizes $\gamma_2$ is 
\beq
\label{eqn:defr2d}
r=(b^2-\delta^2)^{-1}\bigg[ (b^2-a\delta)R +(a-\delta)b^2\frac{\delta}{R} 
\bigg].
\eeq
However, this transformation function is generally not one-to-one.  The problem is illustrated in Figure \ref{fig:fig1}, and comes from the fact that $\dd r/\dd R = 0$ at some value of $R\in (\delta, b)$.   This cannot occur for the mapping function (\ref{eqn:defR2d}).  Consequently equation (\ref{eqn:defR2d}) is a valid transformation for acoustic cloaking while equation (\ref{eqn:defr2d}) is not.

\subsection{Optimal transformation for spherical  cloaks}
We now take \eqref{5=5} as the definition of the global anisotropy measures. 
Using again  the inverse mapping ${\bd X} = f(r) r^{-1}{\bd x}$ it follows  that 
$\nabla^2 {\bd X} = [ (r^2f')'-2f] (fr^2)^{-1} {\bd X}$.  The 
transformation which  minimizes $\gamma_1$ is therefore 
\beq
\label{eqn:defR3d}
f(r)=(b^3-a^3)^{-1}\bigg[ (b^3-a^2\delta)r - (a-\delta)b^3\frac{a^2}{r^2} 
\bigg].  
\eeq
The  transformation which  minimizes $\gamma_2$ is 
\beq
r=(b^3-\delta^3)^{-1}\bigg[ (b^3-a\delta^2)R +(a-\delta)b^3\frac{\delta^2}{R^2} 
\bigg], 
\eeq
but this again has the unphysical nature found for the $\text{2D}$ case.  We conclude that minimization of 
$\gamma_2$ using a single valued function does not appear to have a single or unique solution.  

\subsection{Numerical examples}
The minimizing value of $\gamma_1$ may be found by integrating (\ref{eqn:gamma1}) by parts, and using (\ref{eqn:lapcapX}), 
\beq
\gamma_{1\text{min}} = \begin{cases} 
\frac 1{\pi(b^2-a^2)} \left. \big( 2\pi rf f'\big)\right|_{a}^{b} ,& \text{2D},
\\
\frac 1{\frac43 \pi(b^3-a^3)} \left. \big( 4\pi r^2 f f'\big)\right|_{a}^{b} ,& \text{3D}.
\end{cases}
\eeq
Thus, 
\beq
\label{eqn:gammamin}
\gamma_{1\text{min}} = \begin{cases} 2\, 
(b^2-a^2)^{-2}\big[ (b^2-a\delta)^2 +(a-\delta)^2b^2\big], & \text{2D}, 
\\
3\,  
(b^3-a^3)^{-2}\big[ (b^3-a^2\delta)^2 +(a-\delta)^2b^4\big], & \text{3D}.
\end{cases}
\eeq
\begin{figure}[tb]   
\centering
\subfigure[\label{fig:2drelativegamma} Two   dimensions]
{\includegraphics[totalheight=\figtotalheight]{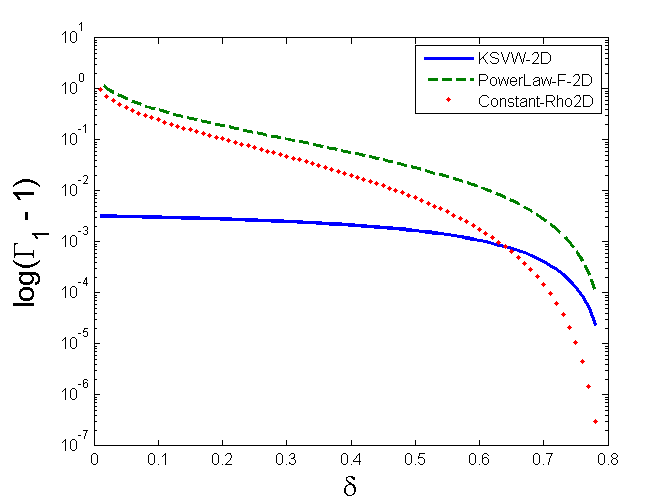}
}
\subfigure[\label{fig:3drelativegamma} Three dimensions]
{\includegraphics[totalheight=\figtotalheight]{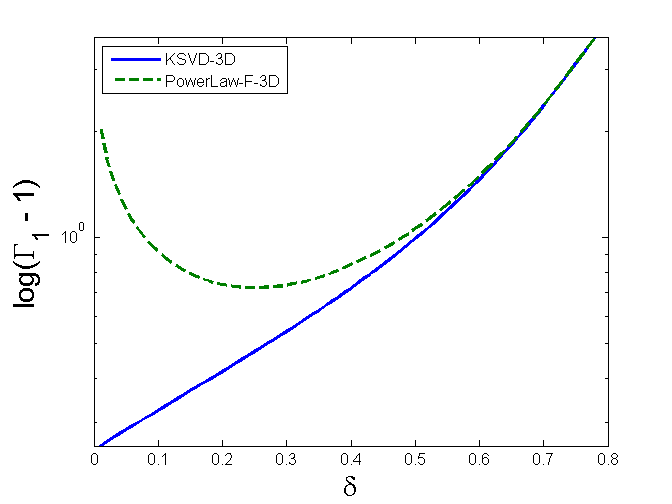}
}
\caption{\label{fig:2d3drelativegamma}The relative value of the global anisotropy parameter for the three transformations defined by equation \eqref{eqn:threetransformations} for $\{a,b\}=\{0.8,1.0\}$.  The curves show $\log (\Gamma_1 -1)$ where  $\Gamma_1 = \gamma_1/\gamma_{1\text{min}}$, with  $\gamma_{1\text{min}}$  given by (\ref{eqn:gammamin}) and $\gamma_{1}$  calculated based on the mappings in equations \eqref{eqn:threetransformations}. 
}
\end{figure}
The relative value of the anisotropy parameter $\gamma_1$ is shown in Figure \ref{fig:2d3drelativegamma} for the three mappings of equation \eqref{eqn:threetransformations}. In all cases, the value of  $\gamma_1$ exceeds the minimum $\gamma_{1\text{min}}$ for the optimal transformations in equations (\ref{eqn:defR2d}) and (\ref{eqn:defR3d}). The KSVW mapping in $\text{2D}$ has anisotropy only slightly more than the minimum, but for $\text{3D}$ the KSVW displays  much larger global anisotropy.

\section{Conclusion and discussion}

Transformation acoustics, like its close analog transformation optics, possesses a huge freedom in the way that the transformation can be chosen.  This paper sheds some light on potential  choices.  We have shown that is possible to always fix at least one of the three material parameters relevant to radially symmetric deformations.  
Starting from the theory of Norris \cite{Norris08b}, we have derived several forms of the transformation $f$ which yield specialized distributions of material properties such as constant and power law density, radial stiffness, and tangential stiffness. This was achieved by reinterpreting the governing equations for the material properties as differential equations for the transformations. We derived a functional form of $f$ that minimizes elastic anisotropy in a certain sense.


\smallskip
\noindent \textbf{Acknowledgement.}
We acknowledge funding from the Office of Naval Research (Contract No. N00014-10-C-260) through their SBIR (Small Business Innovative Research Program) under the supervision of Dr. John Tague and Dr. Jan Lindberg. 

\renewcommand{\bibsection}{\ \ \ \ \ \ \ \-------------}


\end{document}